\documentclass[preprint,12pt]{elsarticle} 
\usepackage{epsfig}
\usepackage{bm}

\begin{document}

\begin{frontmatter}

\title{Dynamic cluster-scaling in DNA}

\author{ A. Bershadskii}

\address{ICAR - P.O. Box 31155, Jerusalem 91000, Israel}

\begin{abstract}
It is shown that the nucleotide sequences in DNA molecules have cluster-scaling properties 
(discovered for the first time in turbulent processes: Sreenivasan and Bershadskii, 2006, 
J. Stat. Phys., 125, 1141-1153.). These properties are relevant to both types of nucleotide pair-bases 
interactions: hydrogen bonds and stacking interactions. It is shown that taking 
into account the cluster-scaling properties can help to improve heterogeneous models of the DNA dynamics. 
Two human genes: BRCA2 and NRXN1, have been considered as examples.     

\end{abstract}

\end{frontmatter}

\section{Introduction}

A DNA molecule carries information in the form of four chemical groups or nucleotide bases: 
adenine, cytosine, guanine, and thymine, represented by the letters A, C, G and T. The order 
of bases on a DNA strand is the DNA sequence. If we read along one of the two DNA-helix sides we get 
text like GATACA... In the double-stranded DNA, the two strands run in opposite directions 
and the bases pair up such that A always pairs with T and G always pairs with C. That is because 
these particular pairs fit exactly to form effective hydrogen bonds with each other. 
The A-T base-pair has 2 hydrogen bonds and the G-C base-pair has 3 hydrogen bonds. 
The G-C interaction is therefore stronger than A-T, 
and A-T rich regions of DNA are more prone to thermal fluctuations and to 
initiation sites (origin) at unwinding stage of DNA replication process. 
The bases are oriented perpendicular to the DNA-helix axis. 
Constant thermal fluctuations result in local twisting, stretching, 
bending, and unwinding of the double-strands. 

In solution DNA assumes linear configuration because it is the one of minimum energy. 
The helix axis of DNA in vivo is usually strongly curved because the stretched length 
of the human genome, for instance, is about 1 meter and this length needs to be "packaged" 
in order to fit in the nucleus of a cell (the diameter of the nucleus from a typical human somatic cell 
is about $5 \times10^{-6}$ meters). Therefore, the DNA has to be highly organized. 
This packaging of DNA deforms it physically, thereby increasing its energy (less stable than relaxed DNA, 
due to less than optimal base stacking). In this situation certain strain is relieved by supercoiling: 
helix bends and twists to achieve better base stacking orientation despite having too many bp/turn. 
The difference in A-T and G-C interactions can be used for optimizing the free energy. 
The base-pairs stacking energies (the main stabilizing factor in the DNA duplex, see for instance 
Yakovchuk et al., 2006) are highly 
dependent on the base sequence (Saenger 1984). These interactions come partly from the overlap 
of the $\pi$ electrons of the bases and partly from hydrophobic interactions. 
Quantum chemistry calculations give rather different
energies for different stacked base pairs: Fig. 1. Therefore, certain clustering of the 
base-pairs can be used by nature in order to minimize the excess energy that builds up when 
DNA molecules are deformed during the process of packaging.

%%%%%%% FIGURE 1 %%%%%%%%%%%%%%%%%%
\begin{figure} \vspace{-3cm}\centering
\epsfig{width=.8\textwidth,file=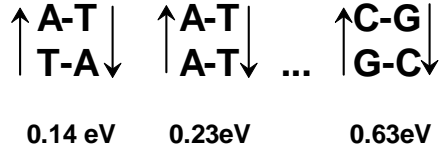} \vspace{-12.5cm}
\caption{The stacking energies for different stacked
base pairs.}
\end{figure}
%%%%%%%%%%%%%%%%%%%%%%%%%%%%%%%%%%% 

Moreover, the increase in stored (potential) energy within  the molecule is then available 
to drive reactions such as the unwinding events that occur during DNA 
replication. Before replication of DNA can occur, the length of the DNA double helix about to be copied must 
be unwound and the two strands must be separated by breaking the hydrogen bonds that link the paired bases. 
The process of replication begins in the DNA molecules at thousands of sites called origins of replication. 
Because the location and time of initiation of origins is generally stochastic, the
time to finish replication will also be a stochastic process. The random distribution of origin firing raises the random gap problem: a random distribution will lead to occasional large gaps that should take 
a long time to replicate. Despite this each cell in a population must complete the
replication process in an accurate and timely manner (see for instance, Hyrien et al., 2003; Jun and Rhind, 2008). 
Different solutions to this problem have been suggested (see, for instance, Blow et al., 2001; Rhind 2006; 
Conti et al., 2007; Yang and Bechhoefer, 2008). 
%%%%%%% FIGURE 2 %%%%%%%%%%%%%%%%%%
\begin{figure} \vspace{-3cm}\centering
\epsfig{width=.8\textwidth,file=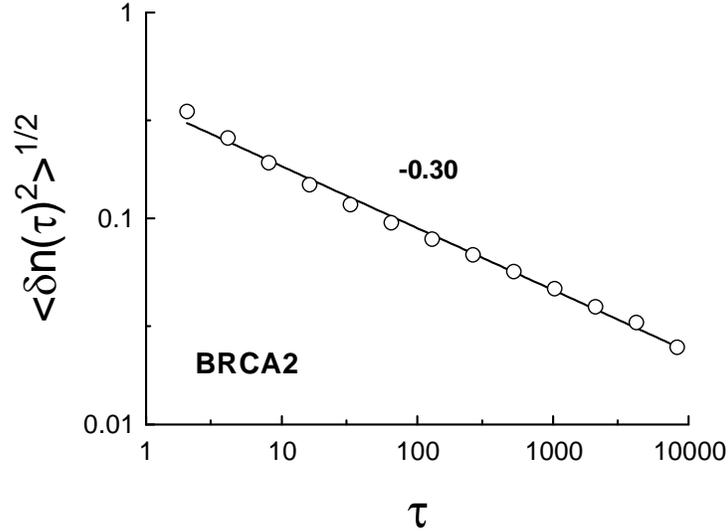} \vspace{-6.5cm}
\caption{The standard deviation for $\delta n (\tau)$ vs $\tau$ for
T-dominated sub-sequence of gene BRCA2 (in log-log scales). The straight
line (the best fit) indicates the scaling law Eq. (2). }
\end{figure}
%%%%%%%%%%%%%%%%%%%%%%%%%%%%%%%%%%% 
If the spacing of origins is not completely random then any regularity in the spacing 
of origins will tend to suppress the large gaps (Blow et al., 2001). For instance, origins within specific 
clusters could be preferred to fire (Mesner et al., 2003; Shechter and Gautier, 2005). Since 
a G-C base pair, with three hydrogen bonds, is expected to be harder to break than an A-T base pair with only 
two bonds, a clustering of these two kinds of the base-pares can be operational in order to solve the random 
origin firing problem. 

\section{Cluster-scaling}

Thus, one can see that certain clustering of the base pairs can be one of nature's solutions for the 
packing and unwinding DNA problems. Because of many orders of space scales involved in these processes 
one can expect that the clustering will exhibit scale-invariant properties (see, for instance, 
Stanley et al., 1999; Bershadskii, 2001). A cluster-scaling for 
stochastic systems was recently suggested in a paper of Sreenivasan and Bershadskii (2006). 
The genome data can be readily checked on the cluster-scaling properties in a "1 or 0" mapping. 
In this presentation (Voss, 1992; Podobnik et al., 2007 and references 
therein) one should put A=1 and C=G=T=0 in an original DNA sequences to obtain an A-dominated sub-sequences 
(one can obtain C or G, or T-dominated sub-sequences in analogous way). Then, to study statistical 
clustering in sub-sequences $\left\{a_i\right\}$ (where $a_i$=1 or 0 and $i=1,2...$) one should take running average:
$$
n_j(\tau) = \frac{1}{\tau} \sum_{i=j}^{i=j+\tau} a_i   \eqno{(1)}
$$   
along the sub-sequences.
For the 1 or 0 mapping this running average will present a weight of the sub-sequences in 
interval [$j$,$j+\tau$]. Following to Sreenivasan and Bershadskii (2006) we are interested in scaling 
variation of the standard deviation of the running density fluctuations $\langle \delta
n_j(\tau)^2 \rangle^{1/2}$ with $\tau$
$$
\langle \delta n_j(\tau)^2 \rangle^{1/2} \sim \tau^{-\alpha}
\eqno{(2)}
$$
where $\langle...\rangle$ denotes average over the sub-sequences, $\delta n_j(\tau)= n_j(\tau) - 
\langle n (\tau) \rangle$. The power law, Eq. (2), corresponds to a scale-invariant (scaling) behavior. 

The exponent $\alpha$ in Eq. (2) was called by Sreenivasan and Bershadskii (2006) as cluster-exponent. 
For white noise zeros (intersections of a white noise signal with time axis) it can be derived analytically that $\alpha = 1/2$ (see Sreenivasan and Bershadskii, 2006). This value can be considered as an upper limit 
(non-clustering case) for the cluster-exponent. If $0< \alpha < 0.5$ we have a cluster-scaling situation, and 
the cluster-scaling is stronger for smaller values of $\alpha$ (see for examples Sreenivasan and Bershadskii, 2006).

%%%%%% FIGURE 3 %%%%%%%%%%%%%%%%%%
\begin{figure} \vspace{-3cm}\centering
\epsfig{width=.8\textwidth,file=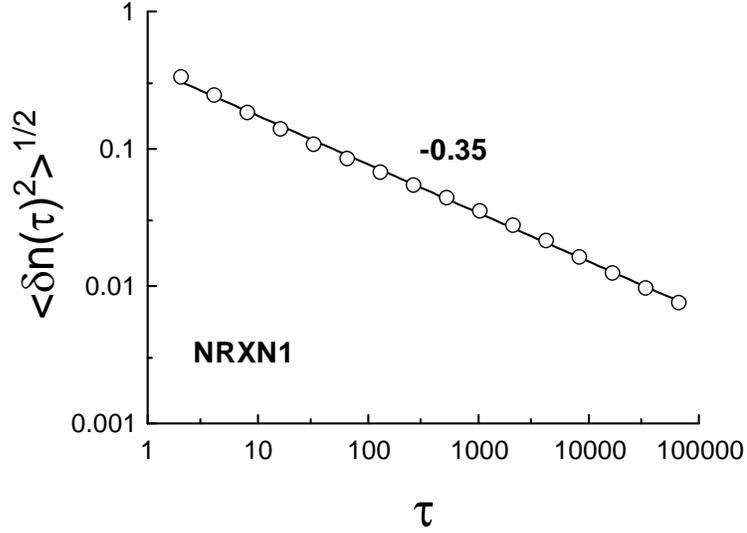} \vspace{-6.5cm}
\caption{The standard deviation for $\delta n (\tau)$ vs $\tau$ for
T-dominated sub-sequence of gene NRXN1 (in log-log scales). The straight
line (the best fit) indicates the scaling law Eq. (2). }
\end{figure}
%%%%%%%%%%%%%%%%%%%%%%%%%%%%%%%%%%% 
In this note we will present, as an example, results obtained for human genome. 
The results of computations for the genome sequences associated with genes: BRCA2 and NRXN1, 
are shown in figures 2 and 3 respectively (the full set of the genome sequences can be found in site: 
http://www.ncbi.nlm.nih.gov). Molecular location of gene BRCA2 on chromosome 13: base pairs 
32,889,616 to 32,973,808). BRCA2 gene helps prevent cells from growing and dividing too rapidly or 
in an uncontrolled way. By helping repair DNA, BRCA2 plays a role in maintaining the stability 
of a cell's genetic information. Gene NRXN1 (neurexin 1) is among the largest known in human,
molecular location on chromosome 2: base pairs 50,145,642 to 51,259,673.
NRXN1 gene represents a strong candidate for involvement in the etiology of nicotine dependence, and even subtle changes in NRXN1 might contribute to susceptibility to autism. \\  

We show in Figs 2 and 3 results for the T-dominated sub-sequences, 
whereas the results for A, C, and G-dominated sub-sequences are similar to those shown in the 
Figs. 2 and 3 (for each gene respectively). Fig. 2 shows (in the log-log scales) dependence of the 
standard deviation of the running density fluctuations $\langle \delta n (\tau)^2 \rangle^{1/2}$ on
$\tau$ for the T-dominated subsequence of gene BRCA2. The straight line is drawn in this figure to indicate the
scaling (2). The slope of this straight line provides us with the cluster-exponent  $\alpha = 0.30 \pm 0.02$. 
Figure 3 shows analogous result for gene NRXN1 with $\alpha = 0.35 \pm 0.02$. One can see that in both cases 
we have rather strong cluster-scaling with the cluster-exponent different for the different genes.  

\section{Hydrogen bonds} 

The most popular potential for modeling the hydrogen (H) bond within a base-pair in the DNA chains is the 
Morse potential (see, for instance, Peyrard and Bishop, 1989; Dauxois et al., 1993; Campa and Giansanti, 1998; 
Hennig and Archilla, 2004):
$$
V_i (y_i) = D_i\left[ \exp-\left(\frac{a}{2} y_i \right) -1\right]^2  \eqno{(3)} 
$$
where $D_i$ is the site-dependent dissociation energy of the $i$th pair, which can take two values $D_i=D^{A-T}$ and 
$D_i= D^{C-G}$ for the A-T and the C-G pairs in the $i$th site respectively (the A-T pair includes two H bonds, 
while the C-G pair includes three H bonds, see Introduction); $a^{-1}$ is a measure of the potential well
width; variable $y_i$ is a dynamical deviation of the H bonds from their equilibrium lengths at position $i$. 
The ratio $D^{C-G}/D^{A-T}=1.5$ is often used for the model 
purposes (though recent quantum chemical calculations by Sponer et al. 2001, results in a ratio 
$D^{C-G}/D^{A-T}=2$).

%%%%%% FIGURE 4 %%%%%%%%%%%%%%%%%%
\begin{figure} \vspace{-3cm}\centering
\epsfig{width=.8\textwidth,file=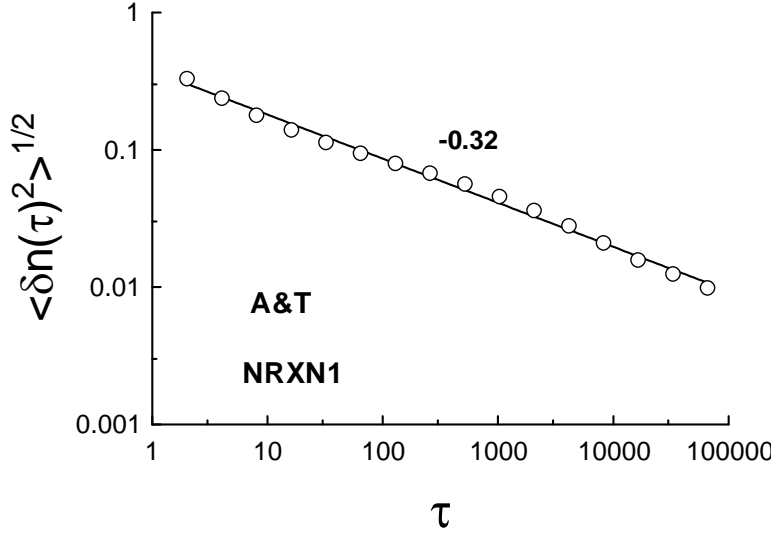} \vspace{-6.5cm}
\caption{The standard deviation for $\delta n (\tau)$ vs $\tau$ for
A\& T (circles) dominated sub-sequence of gene NRXN1
(in log-log scales). The straight lines (the best fit) indicate the scaling law Eq. (2). }
\end{figure}
%%%%%%%%%%%%%%%%%%%%%%%%%%%%%%%%%%% 

Randomly distributed along the DNA chain bivalued H-bond coupling
strengths $D^{A-T}$ and $D^{G-C}$ are usually used in the DNA dynamics models. This would be appropriate 
for an arbitrary base pair sequence. However, as it is follows from previous 
consideration, homogeneous random distribution is not realistic even for the most long genes like 
NRXN1 (see, for instance, Li et al., 1998). 
The dynamic heterogeneous properties of DNA molecule was considered, for instance, as a 
reason for the so-called multi-step melting (Cule and Hwa, 1997). However, the assumption of a random 
and short-range (delta-) correlated sequence made by Cule and Hwa (1997) do not result in the multi-step 
melting and only an additional assumption of an additional backbone stiffness due to the double-stranded 
conformation of DNA molecule allowed to the authors to observe a multi-step melting in their model. 
In the model suggested by Jeon et al. (2007) the sequence randomness considered as a quenched noise with 
finite sequence correlation length. In this approach regions dominated by A-T or, alternatively, by C-G pairs 
play significant role in the bubble (i.e. locally denaturated states) formation. \\

Taking into account the cluster-scaling of the DNA nucleotides is a natural step toward more realistic dynamical 
model. Because of the bivalued H-bond coupling strengths: $D_i=D^{A-T}$ or $D_i=D^{G-C}$, this can be readily 
done using following bivalued mapping: $A=T=1,~ C=G=0$ or, alternatively, $C=G=1,~ A=T=0$. 
Figure 4 shows cluster-scaling behavior, Eq. (2), for the former mapping of NRXN1 gene. The cluster scaling exponent $\alpha = 0.32 \pm 0.02$ in this case. For $C=G=1,~ A=T=0$ the mapping calculations give the same result 
as well as for corresponding mappings of the gene BRCA2 (indication of an universality). Therefore, the bivalued sequences of the $D_i$ coefficients for the DNA dynamic chain should be chosen as cluster-scaling ones with certain 
cluster-exponent $\alpha$ Eq. (2) (for the considered genes $\alpha \simeq 0.32$).       

\section{Stacking interaction}

%%%%%% FIGURE 5 %%%%%%%%%%%%%%%%%%
\begin{figure} \vspace{-3cm}\centering
\epsfig{width=.8\textwidth,file=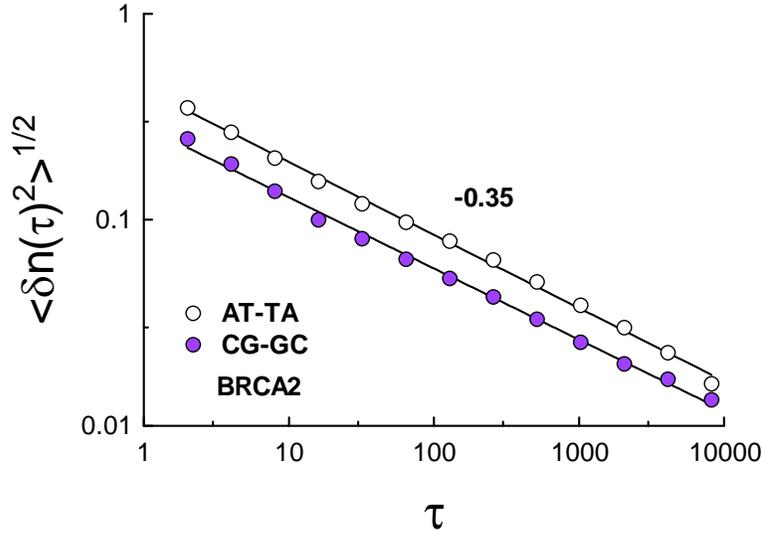} \vspace{-6.5cm}
\caption{The standard deviation for $\delta n (\tau)$ vs $\tau$ for
AT-TA (circles) and CG-GC (filled circles) dominated sub-sequence of gene BRCA2 
(in log-log scales). The straight lines (the best fit) indicate the scaling law Eq. (2). }
\end{figure}
%%%%%%%%%%%%%%%%%%%%%%%%%%%%%%%%%%% 
%%%%%% FIGURE 6 %%%%%%%%%%%%%%%%%%
\begin{figure} \vspace{-3cm}\centering
\epsfig{width=.8\textwidth,file=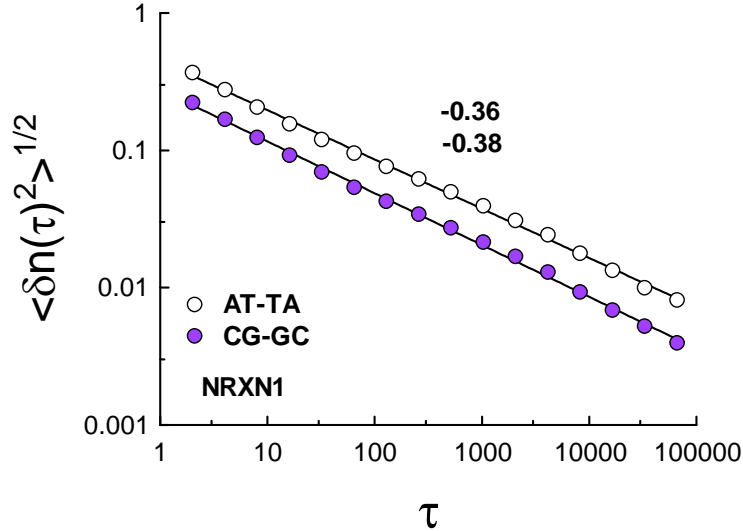} \vspace{-6.5cm}
\caption{The standard deviation for $\delta n (\tau)$ vs $\tau$ for
AT-TA (circles) and CG-GC (filled circles) dominated sub-sequence of gene NRXN1
(in log-log scales). The straight lines (the best fit) indicate the scaling law Eq. (2). }
\end{figure}
%%%%%%%%%%%%%%%%%%%%%%%%%%%%%%%%%%%  

Two factors are mainly responsible for the stability of the DNA double helix: 
base pairing between complementary strands and stacking between adjacent
bases (see Introduction). It is shown experimentally that DNA stability is mainly determined by
base-stacking interactions which contribute greatly into the dependence of the duplex 
stability on its sequence. (see, for instance, a recent paper Yakovchuk et al., 2006). 
Therefore, it is interesting to check whether the staking interactions dominate also the above-considered cluster-scaling phenomenon (cf. Introduction). In order to check this let us use following mapping: in combination $AT=TA=1~1$, 
and $A=T=G=C=0$ otherwise. An alternative mapping is: in combination $CG=GC=1~1$, and $A=T=G=C=0$ 
otherwise. If the stacking interactions dominate the cluster-scaling phenomenon, then one can expect 
that the cluster-scaling will be more pronounced just for these maps (cf. Fig. 1). It means that 
the cluster-exponent corresponding to these maps would be {\it smaller} than cluster-exponents 
observed for the above considered maps. As one can see 
comparing  Figs. 5 and 6 with Figs. 2,3, and 4 in reality we have an opposite situation. 
This comparison indicates that the stacking interactions 
do not dominate the above-considered cluster-scaling phenomenon (at least for the examples given 
in the paper). 

In a realistic dynamic model of DNA molecule one should take into account also the cluster-scaling 
of staking interaction itself as it is shown in Figs. 5 and 6 for instance (see also Ambjornsson et al., 
2006; Krueger et. al., 2006, for heterogeneity of both pairing and stacking interactions). 
This can be done in the frames of a commonly used approximation for the stacking potential (see, for instance, 
Joyeux and Florescu 2009) 
$$
W_i (y_i,y_{i-1}) = \frac{\Delta H_i}{C} \left(1-\exp\left(-b(y_i-y_{i-1})^2\right)\right)  \eqno{(4)}
$$
where $\Delta H_i$ can take different values for different staked pairs $\left\{y_i,y_{i-1}\right\}$. 
Because the situation is not bivalued in this case this task seems to be more difficult than for the hydrogen bonds. 
The main problem here is hybridization of the nucleotides in different types of the stacked base-pairs (Fig. 1). 
The fact that the cluster-scaling exponents for the different types of stacked base-pairs have approximately the same 
value can help to solve this problem. This is not the case, however, for hybridization problem if one will 
consider a realistic model taking into account cluster-scaling of both hydrogen bonds and staking interactions 
(the cluster-scaling exponents are different for hydrogen bonds: Fig. 4, and for staking interactions: Figs. 5 and 6).

\begin{center}

{\bf References}

\end{center} 

Ambjornsson T., Banik S.K., Krichevsky O., and Metzler R., 2006, 
Sequence Sensitivity of Breathing Dynamics in Heteropolymer DNA, 
Phys. Rev. Lett., {\bf 97}, 128105.\\

Bershadskii A., 2001, Multifractal and probabilistic properties of DNA sequences, 
Phys. Lett. A, {\bf 284}, 136-140.\\
  
Blow, J. J., Gillespie, P. J., Francis, D., and Jackson, D. A. 2001, Replication
origins in Xenopus egg extract are 5–15 kilobases apart and are activated in
clusters that fire at different times. J. Cell Biol. {\bf 15}, 15-25.\\

Campa A., and Giansanti A., 1998, Phys. Rev. E, {\bf 58}, 3585-3588.\\

Conti C., Sacc B. Herrick J., Lalou C., Pommier Y., and Bensimon A., 2007, 
Replication fork velocities at adjacent replication origins are 
coordinately modified during DNA replication in human cells, Mol. Biol. Cell., {\bf 18}: 3059-3067.\\

Cule D. and Hwa T., 1997, Denaturation of Heterogeneous DNA, Phys. Rev. Lett., {\bf 79}, 2375-2378. \\   

Dauxois T. and Peyrard M., and Bishop A.R., 1993, Dynamics and thermodynamics of a nonlinear model for 
DNA denaturation, Phys. Rev. E 47, 684-695.\\

Jeon J-H., Park P-J., and Sung W, 2007, The effect of sequence correlation on bubble statistics in double-stranded
DNA, J. Chem. Phys., {\bf 125}, 164901.\\

Joyeux M. and Florescu A-M., 2009, Dynamical versus statistical mesoscopic models for DNA denaturation, 
J. Phys.: Condens. Matter, {\bf 21}, 034101.\\

Jun S. and Rhind N., 2008, Just-in-time DNA replication, Physics, {\bf 1}, 32.\\

Saenger W., 1984, Principles of Nucleic Acid Structure (Berlin: Springer).\\

Hennig D., and Archilla J.F.R., 2004, Multi-site H-bridge breathers in a DNA-shaped double strand. 
Phys. Scr. {\bf 69}, 150-160.\\

Hyrien O., Marheineke K., Goldar A., 2003, Paradoxes of eukaryotic DNA replication: 
MCM proteins and the random completion problem, Bioessays {\bf 25}: 116-125.\\

Krueger A., Protozanova E., and Frank-Kamenetskii M.D., 2006, Biophys. J., {\bf 90}, 3091-3099.\\

Li W., Stolovitzky G., Bernaola-Galvan P., and Oliver J.L., 1998, Compositional Heterogeneity 
within, and Uniformity between, DNA Sequences of Yeast 
Chromosomes, Genome Res., {\bf 8}, 916-928.\\

Mesner, L. D., Li, X., Dijkwel, P. A., and Hamlin, J. L., (2003), The dihydrofolate
reductase origin of replication does not contain any nonredundant nonredundant
genetic elements required for origin activity. Mol. Cell. Biol. {\bf 23},
804-814.\\

Peyrard, M., Bishop, A.R., 1989, Statistical mechanics of a nonlinear model for DNA denaturation. 
Phys. Rev. Lett., {\bf 62}, 2755-2758.\\

Podobnik B., Shaoc J., Dokholyand N.V., Zlatice V., Stanley H.E., and Grosse I., 2007, 
Similarity and dissimilarity in correlations of genomic DNA, 
Physica A, {\bf 373},497-502.\\

Rhind N., (2006), DNA replication timing: Random thoughts about origin firing, Nat. Cell Biol., 
{\bf 8}, 1313-1316.\\

Shechter, D., and Gautier, J. (2005). ATM and ATR check in on origins: a
dynamic model for origin selection and activation. Cell Cycle {\bf 4}, 235-238.\\

Sponer, J., Leszczynski, J. and Hobza, P., 2001, Electronic
properties, hydrogen bonding, stacking, and cation binding of DNA and RNA bases. Biopolymers {\bf 61}, 
3-31.\\

Sreenivasan K.R., and Bershadskii A., 2006, Clustering Properties in Turbulent Signals, 
J. Stat. Phys., {\bf 125}, 1141-1153.\\

Stanley H.E., Buldyreva S.V., Goldbergerb A.L., Havlin S., 
Penga C-K., and Simons M., 1999, Scaling features of noncoding DNA, Physica A, {\bf 273}, 1-18. \\

Voss R.F., 1992, Evolution of long-range fractal correlations and 1/f noise in DNA base sequences, 
Phys. Rev. Lett., {\bf 68}, 3805-3808.\\

Yang S. C-H., and Bechhoefer J., 2008, How Xenopus laevis embryos replicate reliably: 
Investigating the random-completion problem, Phys. Rev. E, {\bf 78}, 041917.\\

Yakovchuk P., Protozanova E., and Frank-Kamenetskii M.D., 2006, Base-stacking and base-pairing contributions into
thermal stability of the DNA double helix, Nucleic Acids Research, {\bf 34}, 564-574.
\end{document}